\documentclass[twocolumn,aps,pra,floatfix,longbibliography]{revtex4-2}

\usepackage{url}	
\usepackage{amsmath}
\usepackage{txfonts}
\usepackage{microtype}
\usepackage{graphicx}
\usepackage{color}
\usepackage{ulem}
\usepackage{bm}
\newcommand{\comm}[1]{}
\usepackage{calrsfs}
\DeclareMathAlphabet{\pazocal}{OMS}{zplm}{m}{n}

\usepackage{mathalfa}

\begin{document}

\title{Nondipole Coulomb effects in strong x-ray field ionization}

\author{Haoyuan He}\email{haoyuan.he@mpi-hd.mpg.de}
\author{Karen Z. Hatsagortsyan}\email{k.hatsagortsyan@mpi-hd.mpg.de}
\author{Christoph H. Keitel}
\affiliation{Max-Planck-Institut f\"ur Kernphysik, Saupfercheckweg 1, 69117 Heidelberg, Germany}

\date{\today}

\begin{abstract}

Atomic ionization in  superintense x-ray fields is studied through numerical solutions of the time-dependent Schr\"odinger equation including nondipole corrections. The role of the Coulomb potential of the atomic core in the nondipole regime is analyzed for a linearly polarized laser field.
In the considered high-frequency and high-intensity stabilization regime, the photoelectron spectra are dominated by a near-zero-energy structure, while above-threshold ionization peaks are suppressed. Two sub-regimes in the nondipole case are identified based on the scaling of the Coulomb momentum transfer, when the Coulomb effect during the interaction with the laser field is either significant or minor. The Coulomb influenced regime sets in
within the realm of relatively low field strengths, high laser frequencies, and long pulse durations. An intuitive model explaining the distinct features in the photoelectron momentum distributions and the Coulomb effect scaling is provided.

\end{abstract}

\maketitle

\section{Introduction}

The dipole approximation has proven successful in the theory of strong-field ionization, leading to the discovery of fundamental nonperturbative phenomena \cite{Becker_2002} and to the development of attoscience \cite{Corkum_2007,Krausz_2009}. Its validity rests on the fact that the typical scale of the dynamics is much smaller than the laser wavelength, which reduces the interaction to that of a spatially uniform, time-dependent electric field. This is especially relevant for the strong-field processes in long wavelength infrared laser fields.
However, at high laser intensities, the magnetic field contribution to the Lorentz force and the spatial variation of the field become non-negligible. This gives rise to nondipole effects, such as the suppression of recollisions \cite{Dammasch_2001,Palaniyappan_2006,Klaiber_2017} and the suppression of high-order harmonic generation \cite{Keitel_1995,Walser_2000a,Milosevic_2000,Kylstra_2001,Kohler_adv}. Recently, advancements in the precision of photoelectron measurement techniques \cite{Weger_2013,Doerner_2000,COLTRIM} have allowed for the observation of subtle nondipole effects -- such as photoelectron momentum shifts along the propagation direction, above-thershold ionization (ATI) peaks shift in opposite direction, the interplay between Coulomb and nondipole effects, \textit{etc} \cite{Klaiber_2005,Klaiber_2013, Ludwig_2014,Maurer_2018,Willenberg_2019,Hartung_2019,Haram_2019,Grundmann_2020,Lin_2021,
Hartung_2021,Lin_2022a,Mao_2025} -- even at relatively weak, nonrelativistic intensities of infrared lasers.

Another type of nondipole effects in ionization can be observed in short wavelength laser fields,  when the laser wavelength is comparable with the atomic size. Photoionization in hard x-ray fields shows this type of nondipole effect in the  single photon perturbative regime \cite{Sommerfeld_1930,Schmidt_2024}. 
With the advent of powerful free-electron laser facilities, such as the European x-ray free-electron lasers (XFEL) in Hamburg, the Linac Coherent Light Source at Stanford,  China’s upcoming soft x-ray free-electron lasers in Shanghai and others, strong laser fields have been reachable in high-frequency domain, which allowed for the observation of nonlinear ionization processes with extreme-ultraviolet (XUV) and x-ray light \cite{Rohringer_2012,Young_2018,Linker_2025}. The upgrades of these facilities promise more intense x-rays with shorter wavelengths. This may allow to enter nonperturbative and nondipole regimes of x-ray atom interaction. The nondipole effects arising  from the magnetic-field-induced Lorentz force becomes important when the laser  strong field parameter $a_0=v_E/c=E_0/(c\omega)$ is nonnegligible, with the electron oscillatory velocity $v_E$ in the laser field, the laser field amplitude $E_0$, the frequency $\omega$, and the speed of light $c$. Atomic units are used throughout, unless stated otherwise. In XFEL fields the regime of $a_0\sim 0.1$ could be reachable \cite{XFEL,FLASH}, where the nondipole signatures are already visible.

Nondipole electron dynamics in strong-field ionization driven by linearly polarized laser high-frequency fields is strongly shaped by the Coulomb potential of the parent ion. The previous studies of this regime \cite{Forre_2006,Zhou_2013,Telnov_2020,Telnov_Chu_2021,Geng_2021} revealed near-zero-energy structures (ZES) associated with nondipole effects. This structure in the form of three-lobe (or multi-lobe) pattern in the photoelectron momentum distribution (PMD) arises from the interplay between the Lorentz-force–induced drift along the laser propagation direction and the Coulomb attraction of the parent ion, which together break the forward–backward symmetry predicted by the dipole approximation. The resulting three-lobe pattern, counterintuitively oriented opposite to the laser propagation direction, provides clear evidence of deviations from dipole-based descriptions in strong-field ionization. As the laser intensity increases, the Lorentz force drives the electron wave packet further away from the nucleus, while the subsequent spreading and the Fresnel-type diffraction of the wave packet on the atomic core leads to a petal-like interference structure opposite to the propagation direction \cite{Geng_2021}.

In this work, we investigate in detail the role of the Coulomb potential in shaping nondipole effects in superintense x-ray laser fields with linear  polarization by solving the time-dependent Schr\"odinger equation (TDSE) including nondipole corrections. We analyze the Coulomb field effect during the interaction with the laser field and found essential differences in the scaling of the Coulomb momentum transfer (CMT) with respect to the laser strong field parameter $a_0$ at strong and weak fields. Based on these results, we define sub-regimes of the nondipole interaction regime, and investigate further the  pulse duration effect for these sub-regimes. We have established that the role of the Coulomb field during the interaction  is essential  at low field strengths, high frequencies, and long pulse durations, with the typical pronounced signatures in ZES.

The structure of the paper is the following. In Sec.~\ref{sec:theo} the theoretical methodology is introduced. The results of simulations of the photoelectron momentum distributions (PMDs)  are discussed in Sec.~\ref{sec:results}. Classification of the nondipole regimes into the  Coulomb-influenced vs. laser-dominated one is given in Sec.~\ref{sec:sub-regimes}. Pulse-duration dependence across these regimes is discussed in Sec.~\ref{sec:pulse}. Our conclusions are given in Sec.~\ref{sec:concl}.

\section{THEORETICAL METHODOLOGY}\label{sec:theo}

We investigate the nondipole regime of strong field ionization in intense x-ray fields, when the field parameter is small, $a_0\lesssim 0.3$, but not negligible. In this case, the dynamics is described by the TDSE with nondipole corrections
\begin{equation}
    i\,\frac{\partial}{\partial t}\,|\Psi(t)\rangle
    = \hat{H}\,|\Psi(t)\rangle ,
\end{equation}
with the Hamiltonian
\begin{eqnarray}
\hat{H}=\frac{1}{2}\left[\hat{\mathbf{p}}+\mathbf{A}(\phi) \right]^2+ V(\mathbf{r}),
\end{eqnarray}
where  $\mathbf{A}(\phi)$ is the  vector potential of the laser field propagating along the $z$ axis and polarized in the $(x,y)$ plane, with the phase $\phi=\omega(t-z/c)$, and $V(\mathbf{r})$ is the atomic potential. To incorporate leading-order nondipole effects $\sim 1/c$, we consider the first-order nondipole expansion of the Hamiltonian in the velocity gauge.  The resulting Hamiltonian reads
\begin{align}
\label{HHH}
    \hat{H}(t)
    &= -\frac{1}{2}\nabla^{2}
    + V(\mathbf{r})
    \underbrace{- i\,A_{x}(t)\frac{\partial}{\partial x}
    - i\,A_{y}(t)\frac{\partial}{\partial y}}_{\hat{H}_{\mathrm{Dipole}}}  
    \nonumber \\
    &\quad 
    \underbrace{+\frac{z}{c}\left[
        A_{x}(t)E_{x}(t)
        + A_{y}(t)E_{y}(t)
    \right]}_{\hat{H}_{\mathrm{Nondipole}}^{(1)}} 
       + \underbrace{- i\,\frac{z}{c}\left[
        E_{x}(t)\frac{\partial}{\partial x}
        + E_{y}(t)\frac{\partial}{\partial y}
    \right]}_{\hat{H}_{\mathrm{Nondipole}}^{(2)}}     ,
\end{align}
where  $\mathbf{A}(t)=(A_x(t),A_y(t))$ is the dipole-approximated vector potential of the laser field, with
the corresponding electric field  $\mathbf{E}(t)= -\partial_t \mathbf{A}(t)$.
The purely time-dependent term $A^2(t)/2$, is removed via a gauge transformation.
 The nondipole correction to the Hamiltonian  is given by two terms:  $\hat{H}_{\mathrm{Nondipole}}^{(1)}$ and $\hat{H}_{\mathrm{Nondipole}}^{(2)}$. They effectively describe the laser magnetic field induced drift-force $v\times B$ along the propagation direction, where the transverse velocity in the drift-force  is due to the laser field in the term $\hat{H}_{\mathrm{Nondipole}}^{(1)}$, and  the term $\hat{H}_{\mathrm{Nondipole}}^{(2)}$ accounts for the effect of the additional transverse momentum.

We use a laser pulse of the following form:
\begin{eqnarray}
    A_x(t)
    &= \frac{E_0}{\omega}\,
       \sin^2\!\left( \frac{\omega t}{2 N_c} \right)
       \sin(\omega t),  \quad
    A_{y,z}(t)     = 0,
            \label{pulse}
\end{eqnarray}
where $E_0$ is the peak amplitude of the electric field, and $N_c$ is the number of optical cycles defining the pulse duration.

To solve the TDSE, the wave function $\Psi(\mathbf{r},t)$ is expanded in spherical harmonics $Y_{\ell}^{m}(\Omega)$ as
\begin{equation}
\Psi(\mathbf{r},t) = \frac{1}{r}
\sum_{\ell=0}^{\infty}
\sum_{m=-\ell}^{\ell}
\Phi_{\ell m}(r,t)
Y_{\ell}^{m}(\Omega),
\end{equation}
where $\Phi_{\ell m}(r,t)$ are the radial wavefunction components associated
with each angular momentum channel $(\ell, m)$, and
$\Omega$ is the solid angle.
%defined by $\mathrm{d}\Omega_r = \sin\theta_r\, \mathrm{d}\theta_r\, \mathrm{d}\varphi_r$. 
For the numerical solution of TDSE with nondipole corrections, we have used Qprop-ND code \cite{Qprop-ND} by extending  the standard Qprop code \cite{Bauer_2006} in the nondipole regime, adding first-order nondipole corrections. Each radial component of the wave function is propagated in real time using an implicit Crank-Nicolson scheme with absorbing boundary conditions. The atomic bound state is obtained by the imaginary-time propagation of the field-free Hamiltonian until the ground-state energy is fully converged.

For the calculation of PMD, the iSURF method is applied \cite{Morales_2016,Tulsky_2020}, which necessitates to replace in the simulation the atomic binding potential $V(\mathbf{r})$ by
\begin{equation}
U(r) =
\begin{cases}
    -\dfrac{1}{r}, & r < R_{\mathrm{co}}, \\[8pt]
    -\dfrac{2R_{\mathrm{co}} - r}{R_{\mathrm{co}}^{2}}, & R_{\mathrm{co}} < r < 2R_{\mathrm{co}}, \\[8pt]
    0, & r > 2R_{\mathrm{co}} ,
\end{cases}
\end{equation}
where the Coulomb potential is switched to a linear form after the cutoff radius $R_{\mathrm{co}}$ (in the present calculations  $R_{\mathrm{co}} = 100$), and then is turned-off. 
 
\section{Simulation results}\label{sec:results}

\subsection{Domination of the zero-energy structure}\label{ZES}

We investigate the nondipole regime of strong-field ionization of atomic hydrogen driven by high-frequency laser fields. The laser frequencies considered are $\omega = 3$~a.u. ($\approx 82$~eV) and $\omega = 5$~a.u. ($\approx 137$~eV), as representatives  of the XUV and soft-x-ray spectral domains. Both photon energies are experimentally available at free-electron laser facilities such as FLASH at DESY, whose FEL photon-energy range extends well beyond this energy region \cite{FLASH}. Particular attention is paid to the role of the Coulomb potential in shaping nondipole effects in the high-frequency strong field regime. In the present parameter range, the ionization yield is strongly concentrated in the near-zero-energy region \cite{Forre_2006,Zhou_2013,Telnov_2020,Telnov_Chu_2021,Geng_2021,Azizi_2025}. Consequently, our analysis focuses on the formation and evolution of nondipole features of the ZES.

\begin{figure}
    \centering
    \includegraphics[width=1\linewidth]{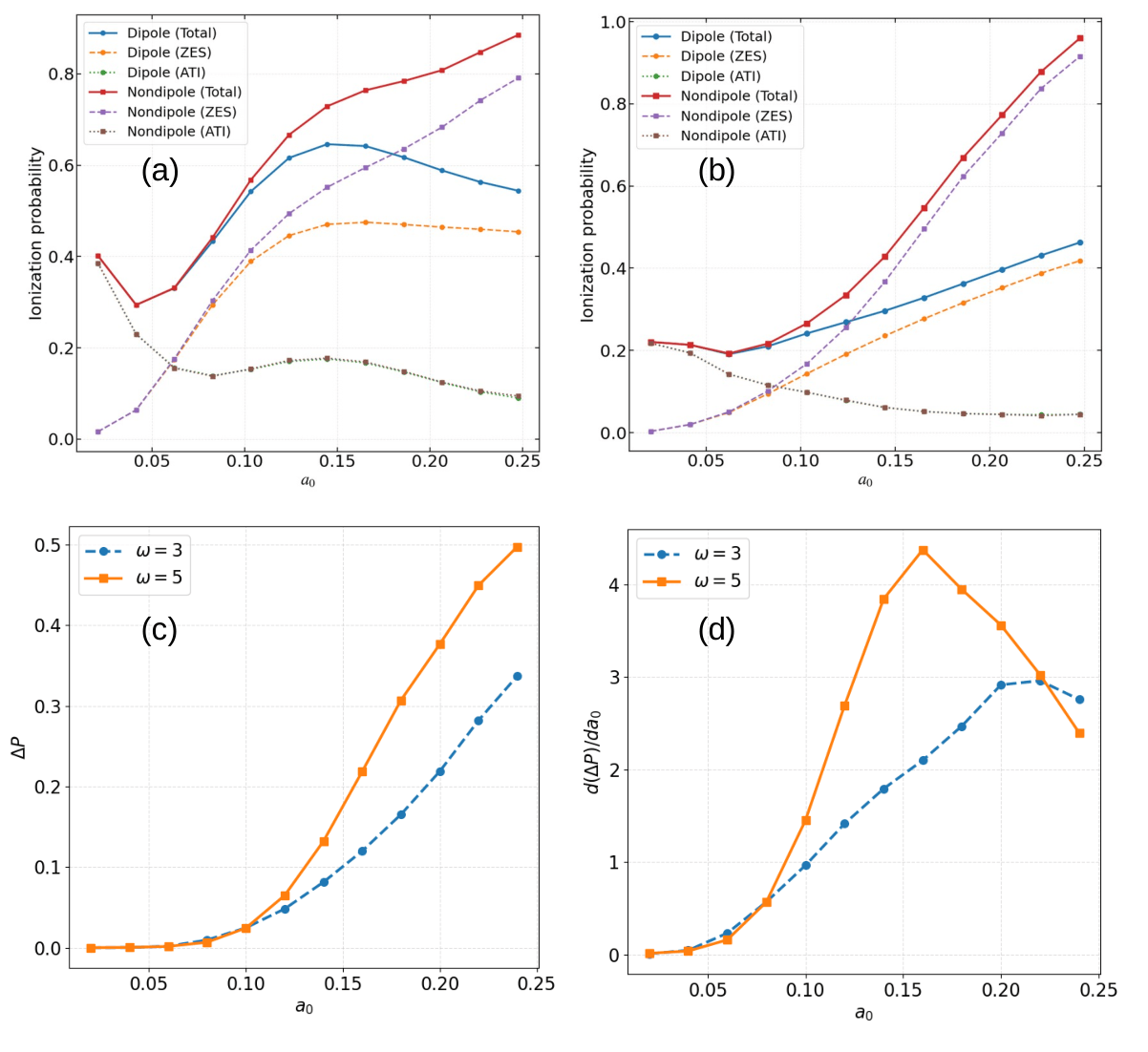}
    \caption{Ionization probabilities vs the laser strong-field parameter $a_0$: (a) for $\omega=3$~a.u. , (b)  $\omega=5$~a.u. Here the total ionization probability together with the ZES ($\varepsilon <\omega/2$) and ATI ($\varepsilon>\omega/2$) contributions are calculated within the dipole and nondipole approximations. (c) Nondipole ionization correction, $\Delta P=P_{\rm ND}-P_{\rm D}$. (d) First derivative of the ionization correction with respect to $a_0$, $d(\Delta P)/da_0$.}
    \label{fig1}
\end{figure}

Our calculations confirm the domination of ZES in PMD. Thus, in Figs.~\ref{fig1}(a,b) we show the total ionization probabilities  as a function of the strong-field parameter~$a_0$, together with their decomposition into the ZES (defined here as the photoelectron energy $\varepsilon<\omega/2$) and ATI structure ($\varepsilon>\omega/2$), for $\omega=3$ and $5$~a.u., respectively, keeping the total pulse duration fixed $\tau=4\pi$  in both frequency cases. We compare also the results of calculations within the dipole and nondipole approximations in Figs.~\ref{fig1}(c,d).

\begin{figure*}
    \centering
    \includegraphics[width=1\linewidth]{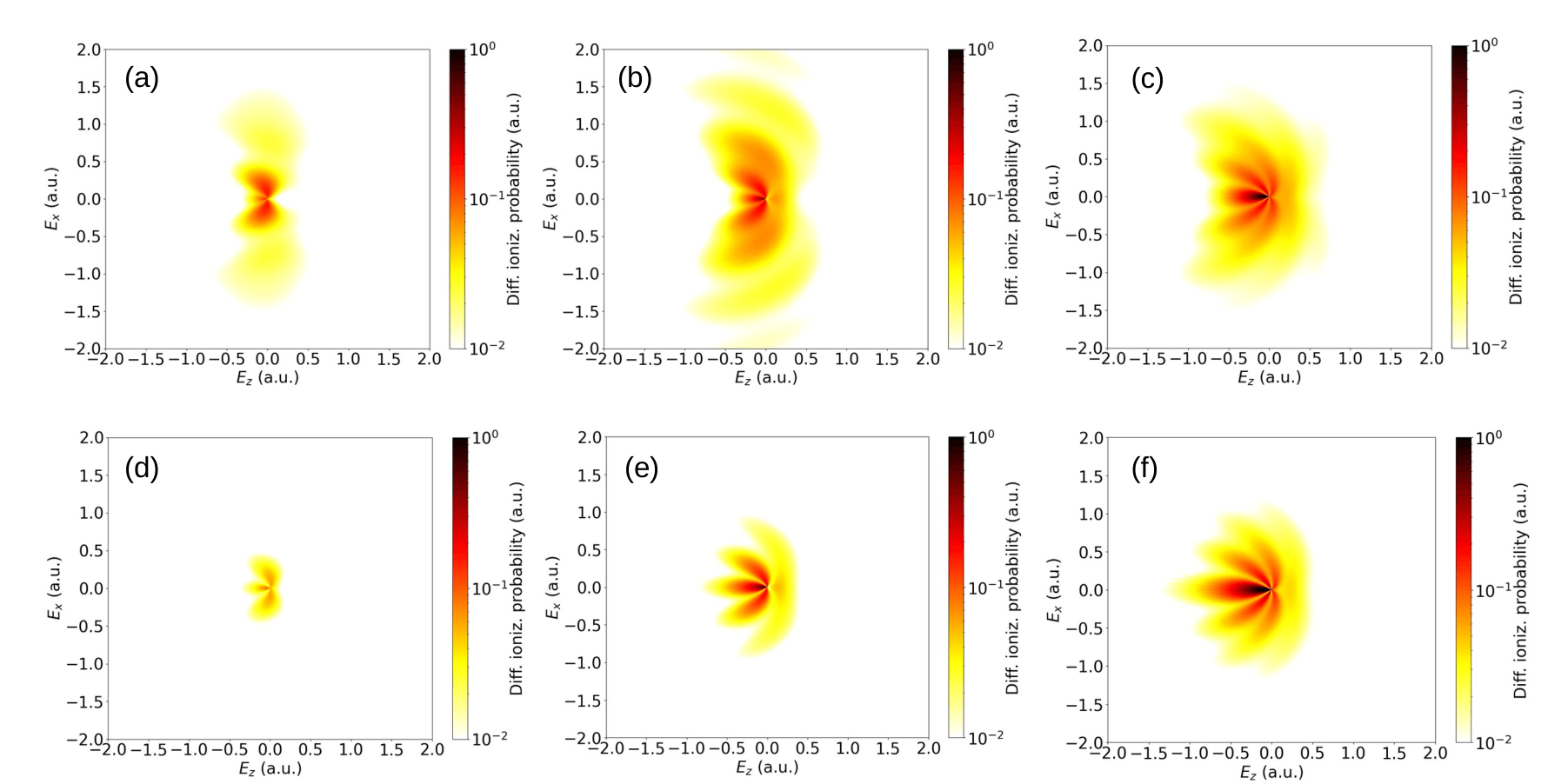}
    \caption{Zero-energy structures (ZES) for a hydrogen atom interacting with a high-frequency linearly polarized laser pulse of duration $\tau=4\pi$. The laser frequencies are $\omega=3$~a.u. (top row), $\omega=5$~a.u. (bottom row). The normalized vector potentials are $a_0=0.083$ (left column), $a_0=0.14$ (middle column), and $a_0=0.206$ (right column), corresponding to the $a_0$-regions before, around, and after the $d(\Delta P)/da_0$-peak shown in Fig.~\ref{fig1}(d). Here, $\varepsilon_x=\varepsilon \sin\theta=pp_x/2$, and $\varepsilon_z=\varepsilon \cos\theta=pp_z/2$, with the photoelectron energy $\varepsilon =p^2/2$, momentum module $p$, and the momentum components $p_{x,z}$. The laser pulse duration  is fixed at $\tau=4\pi$ in all calculations. }
    \label{fig2}
\end{figure*}

Up to $a_0\lesssim 0.07$ the ionization is in the dipole regime, as the dipole and nondipole calculations coincide, see  Figs.~\ref{fig1}(c) for the ionization probability difference between the dipole and nondipole calculations, $\Delta P=P_{\rm ND}-P_{\rm D}$. In the nondipole regime the ATI is dominating over ZES [Figs.~\ref{fig1}(a,b)]. For larger $a_0$, the ZES probability is significantly increased, while the total ATI probability remains nearly constant, even slightly decreasing. Thus, in the nondipole regime ZES dominates over ATI, for instance, at $a_0=0.2$ the probability ratio of ZES to ATI is about 5 for $\omega=3$ (or $\sim 10$ for $\omega=5$).  Further, the common feature at both driving frequencies is that the nondipole enhancement of the total ionization probability originates almost entirely from the ZES, whereas the ATI contribution remains nearly identical in the dipole and nondipole calculations throughout the investigated field range. This observation indicates that the nondipole Lorentz force primarily modifies the ionization yield of low-energy electrons that remain under the influence of the ionic Coulomb potential for an extended period, while its effect on the yield of high-energy electrons is comparatively weak.

The ZES total probability is monotonously increasing with increasing $a_0$, however this dependence in not uniform. For further insight into the evolution of the nondipole correction to the ionization yield, we calculate the first derivative of $\Delta P$ with respect to the strong-field parameter, $d(\Delta P)/da_0$, see Fig.~\ref{fig1}(d). Since at weak fields the ATI exhibits only negligible differences between the dipole and nondipole calculations, the nondipole ionization correction remains very small in this regime. As a result, $\Delta P$ increases only slowly with the field strength [Fig.~\ref{fig1}(d)]. For both driving frequencies, the derivative shows non-monotonous behavior: it initially increases with $a_0$, reaches a well-defined maximum, and subsequently decreases at higher field strengths. We underline that this non-monotonous behavior hints for the qualitative change of the underlying physical dynamics, and  the peak provides a marker for the crossover between possible sub-regimes.  The field strength of the crossover depends on the driving frequencies. For $\omega=5$~a.u., it is lower ($a_0\approx 0.16$) than for $\omega=3$~a.u.  ($a_0\approx0.22$). In next sections we will analyze  the dynamical picture behind the given sub-regimes.

Beyond the crossover, the increase of $\Delta P$ gradually slows down despite the continuing rise of the total ionization probability. This behavior indicates that the nondipole yield enhancement tends to saturation as the electron is displaced progressively farther from the ionic core at large $a_0$.

\subsection{Photoelectron momentum distributions }

When investigating the laser intensity dependence of the ionization yield in the previous section, we have been hinted [Fig.~\ref{fig1}(d)] that the nondipole regime is divided into two sub-regimes  around intermediate $a_0\sim 0.2$. To get more detailed information on the sub-regimes, we calculate PMD depending on $a_0$. The two-dimensional PMDs for ZES in the laser propagation--polarization plane are presented in Fig.~\ref{fig2} for three representative values of the normalized vector potential, $a_0 = 0.083$, $0.14$, and $0.206$, and for laser frequencies $\omega=3$ and $5$~a.u. .

Generally, the PMD exhibits a complex structure characterized by the side lobes  and the central emission main-lobe in the counterpropagation direction. The transition from a side-lobe-dominated to a main-lobe-dominated distribution is highly sensitive to both the laser strong field parameter $a_0$ and the driving frequency $\omega$. The transition from the low $a_0$ [Fig.~\ref{fig2}(a,d)] to the high $a_0$   Fig.~\ref{fig2}(c,f)] is characterized by the main-lobe increase and enhancement of side lobes. The main-lobe increase is significant at higher frequency  $\omega=5$~a.u., while the side lobes enhancement is prominent for low frequency $\omega=3$~a.u. As a result,  at the lower frequency, $\omega=3$~a.u. [Figs.~\ref{fig2}(a-c)], the PMD  is characterized by pronounced side lobes whose intensity and spatial extent significantly exceed those of the central emission lobe. As the laser frequency increases to $\omega=5$~a.u. [Figs.~\ref{fig2}(d-f)], the PMDs' qualitative structure is modified: the side lobes are progressively suppressed, while the central lobe becomes increasingly dominant.

For a more quantitative analysis of PMDs,  we show in Figs.~\ref{fig3}(a) the maximum energy of the main photoelectron lobe in the laser counterpropagation direction. The maximum photoelectron energy $\varepsilon_z$ in the main PMD lobe increases monotonically with increasing $a_0$ [Fig.~\ref{fig3}(a)], more slowly for $\omega=3$, and faster for $\omega=5$. However, this increase is not uniform. The non-uniformness is characterized by $d\varepsilon_z/da_0$ shown in Fig.~\ref{fig3}(b). For both driving frequencies, $d\varepsilon_z/da_0$ shows  a well-defined maximum with respect to $a_0$, similar to Fig.~\ref{fig1}(d), and hints again   that the physical picture of the interaction is different in high and low $a_0$ domain. The border of these domains  determined by the maximum of $d\varepsilon_z/da_0$, is similar to that determined from $d(\Delta P)/da_0$ in Fig.~\ref{fig1}(d). 
  
\section{Coulomb-influenced vs laser-dominated dynamics}\label{sec:sub-regimes}

\begin{figure}
    \centering
    \includegraphics[width=1\linewidth]{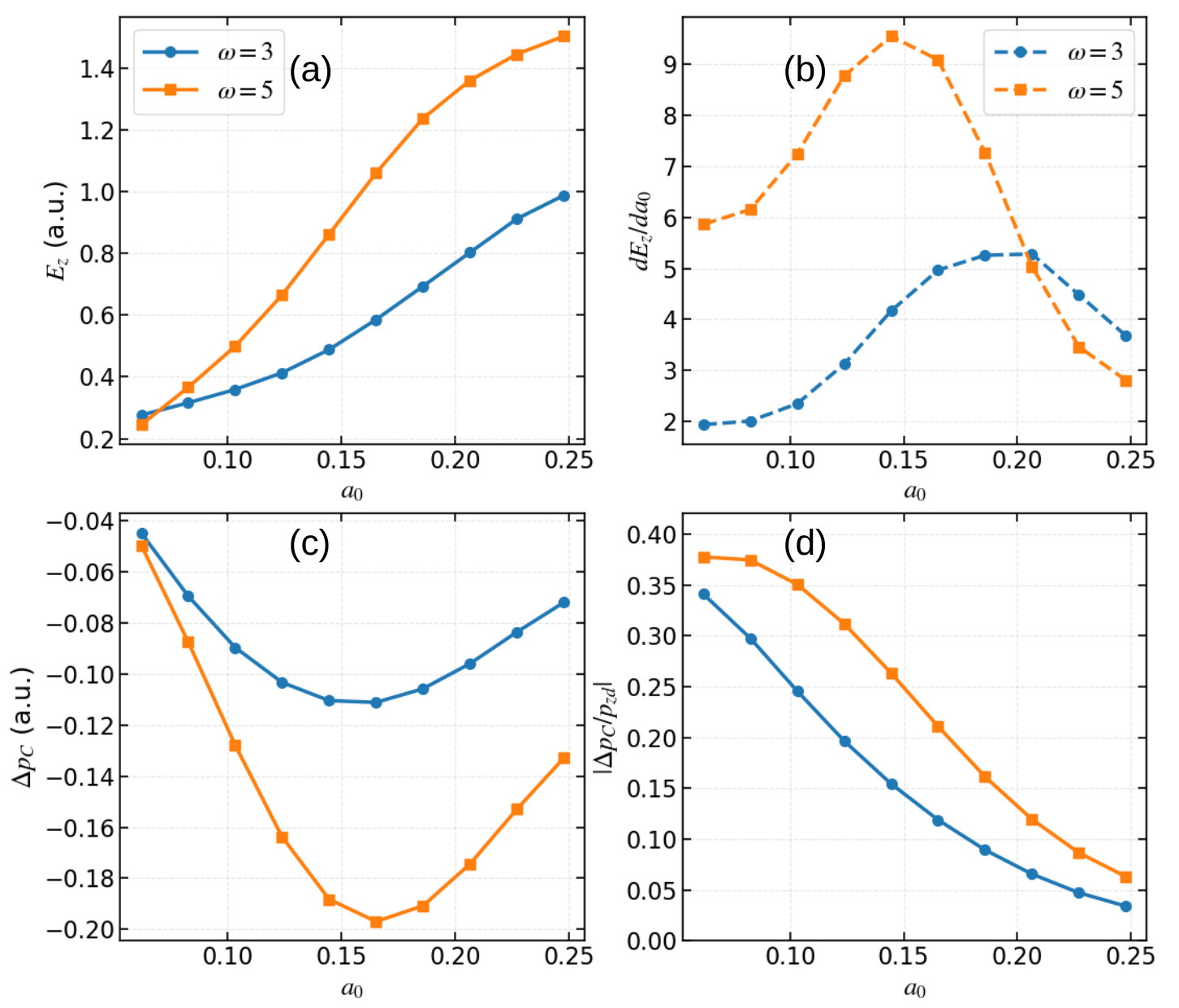}
    \caption{(a) Maximum energy of the main photoelectron lobe as a function of the strong-field parameter $a_0$. (b) First derivative of the maximum photoelectron energy with respect to $a_0$. (c) Accumulated Coulomb momentum transfer (CMT) along the propagation direction at the end of the laser pulse as a function of $a_0$. (d) Ratio of the Coulomb momentum transfer to the laser-induced drift momentum, $\Delta p_{\mathrm{C}}/p_{z\mathrm{d}}$, where $p_{z\mathrm{d}}\sim ca_0^2/4$ is the laser induced drift momentum.}
    \label{fig3}
\end{figure}

Continuing our scrutiny of the origin of the two sub-regimes discussed above, we calculate the accumulated  CMT to the ionized electron in the propagation direction during the interaction, see Fig.~\ref{fig3}(c):
\begin{equation}
\Delta p_{C}=-\int_{-\infty}^{\infty}\left\langle\partial_zV(\mathbf r)\right\rangle dt'.
\end{equation}
It is known \cite{Forre_2006} that the counterintuitive lobe in ZES arises due to the CMT in the propagation direction and it provides the most evident measure of this evolution. Usually, larger is the longitudinal CMT, larger will be the probability of the counterintuitive lobe, as the CMT enables energy absorption from the laser field and facilitates the ionization.

We see from Fig.~\ref{fig3}(c) that the magnitude of the CMT, $|\Delta p_{C}|$, initially increases with $a_0$ at low field strengths, while this trend is reversed at larger $a_0$. The maximum of $|\Delta p_{C}|$ occurs at around $a_0\approx0.165$ for both frequencies. This characteristic value is close to, but does not exactly coincide with the crossover identified from the main-lobe energy in Fig.~\ref{fig3}(b). In particular, the crossover inferred from the energy response occurs at a larger $a_0$ for $\omega=3$ ($a_0\gtrsim0.186$) than for $\omega=5$ ($a_0\gtrsim0.145$).

The CMT also exhibits a pronounced frequency dependence, with a larger $|\Delta p_{C}|$ at the higher frequency. This behavior can be understood from the frequency dependence of the quiver amplitude along the laser polarization direction: $\alpha_0=\ E_0/\omega^2=c a_0/\omega$.
At fixed $a_0$, the excursion in the polarization direction is larger at lower frequencies. At $\omega=3$, the larger transverse excursion drives the electron farther away from the ionic core in the transverse direction, thereby reducing the Coulomb interaction and resulting in a smaller CMT than for $\omega=5$ [Fig.~\ref{fig3}(c)], and consequently, in the lower main-lobe energy [Fig.~\ref{fig3}(a)].

\begin{figure}
    \centering
    \includegraphics[width=1\linewidth]{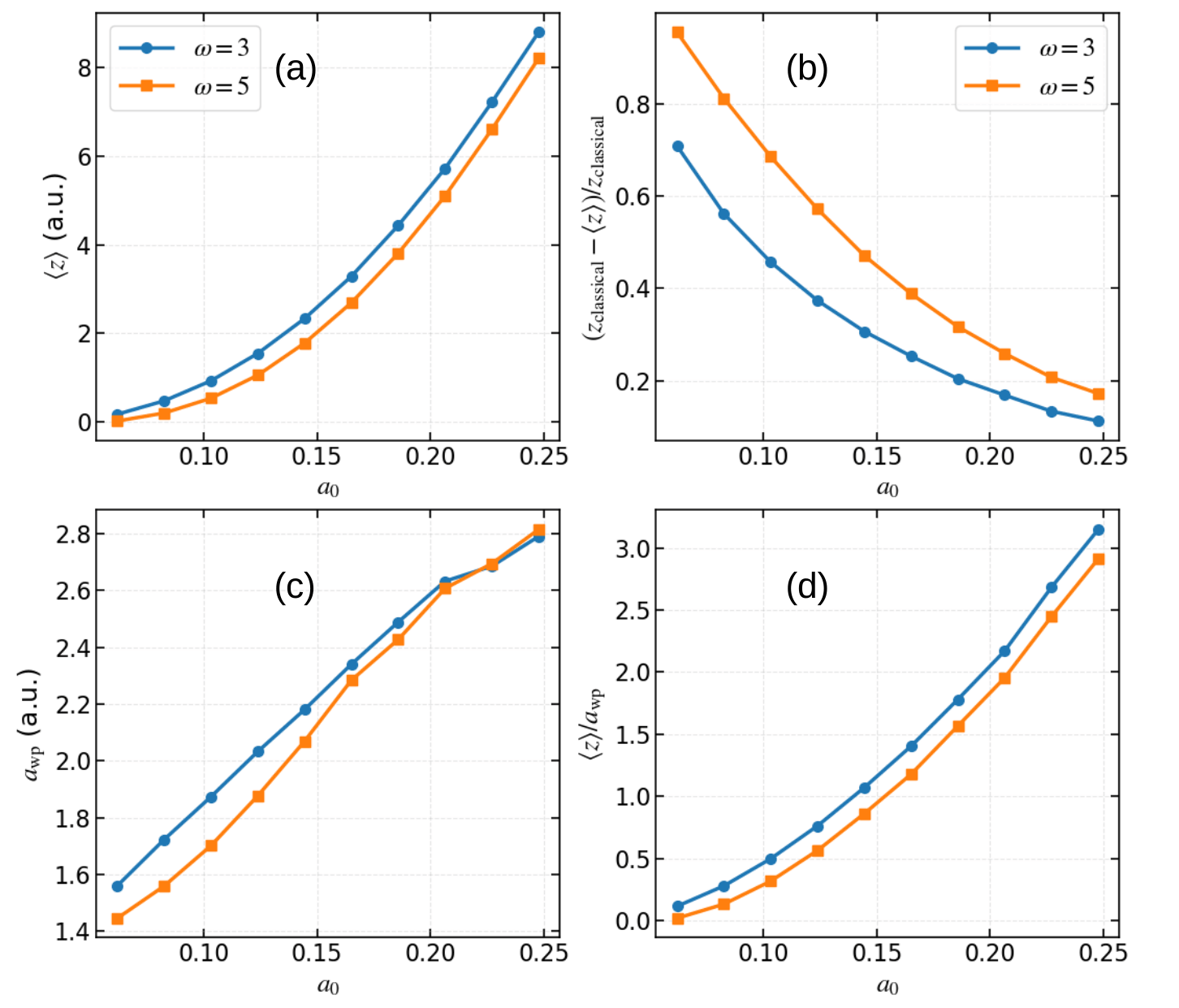}
    \caption{(a) Quantum expectation value of the longitudinal electron displacement, $\langle z\rangle$, at the end of the laser pulse as a function of $a_0$. (b) Relative deviation between the classical and quantum longitudinal displacements, $|z_{\mathrm{cl}}-\langle z\rangle|/z_{\mathrm{cl}}$, as a function of $a_0$. (c) Longitudinal wave-packet size, $a_{\rm wp}$, at the end of the laser pulse as a function of $a_0$. (d) Ratio of the longitudinal electron displacement to the wave-packet size, $\langle z\rangle/a_{\rm wp}$, as a function of $a_0$. }
    \label{fig4}
\end{figure}

The crossover value for $a_0$ between the two different behaviors with respect to the Coulomb effect is slightly different when extracted from the CMT [Fig.~\ref{fig3}(c)], or from the main lobe energy [Fig.~\ref{fig3}(b)]. The reason is that the main lobe energy does not depend solely on the CMT, but also on the transverse spreading of the electron wave packet, quantified by the quiver amplitude $\alpha_0$. The spreading is larger for $\omega=3$ in Fig.~\ref{fig3}(b) which results in a larger shift of the crossover value for $a_0$. The spreading effect is also the reason for the enhanced side lobes of ZES at $\omega=3$ in Fig.~\ref{fig2}(a-c).

A complementary measure of the relative importance of the Coulomb interaction is provided by the ratio of the accumulated CMT to the laser-induced longitudinal drift momentum, $p_{z\rm d}\sim c a_0^2/4$, see  Fig.~\ref{fig3}(d). Unlike the absolute CMT, the absolute value of this ratio decreases systematically with increasing $a_0$ for both frequencies. At weak fields $\Delta p_{\mathrm{C}}/p_{z\mathrm{d}}$ is almost constant, while at large $a_0$ it is sharply tends to zero. This behavior is also indicative of the dynamical transition at intermediate $a_0$. At weak fields ($a_0\rightarrow 0$) the Coulomb effect (CMT) is more important than at high fields ($a_0\rightarrow 1$). The larger ratio observed for $\omega=5$ further indicates that the Coulomb interaction remains relatively more important at a given $a_0$ for the higher-frequency field, pointed out above.

Why does the CMT $|\Delta p_{C}|$ increase with $a_0$ at small $a_0$ and decrease at larger values? At first glance, increasing $a_0$ enhances the laser-induced drift and consequently increases the distance between the ionized electron and the atomic core, which should lead to a reduction in the CMT. This picture is indeed appropriate on the high-field side, where the electron is rapidly displaced from the core and the Coulomb interaction is correspondingly suppressed. At lower field strengths, however, the electron cannot be regarded as a point particle. Instead, the ionized electron is described by an extended wave packet. When the wave-packet size $a_{\rm wp}$ is comparable to or larger than the effective drift distance of the wave packet center $z_0$, different parts of the wave packet can remain on opposite sides of the ionic core and generate partially compensating CMTs. Increasing $a_0$ then enhances the displacement of the wave packet and progressively breaks this cancellation, resulting in an increase of $|\Delta p_{C}|$. Once the drift distance becomes sufficiently large compared with the spatial extent of the wave packet, the point-particle picture becomes more appropriate and the increasing separation from the core leads to a decrease of the accumulated CMT.

To confirm the picture outlined above, we have quantified the wave-packet size  directly from the simulations by the longitudinal spatial variance of the wave packet,
\begin{equation}
a_{\rm wp}\sim
\sqrt{\langle z^2\rangle-\langle z\rangle^2}/2,
\end{equation}
and compare this effective longitudinal width with the quantum longitudinal displacement $\langle z\rangle$, see Fig.~\ref{fig4}.
While the longitudinal displacement $\langle z\rangle$ increases approximately quadratically with $a_0$, following the scaling of the laser-induced longitudinal drift [Fig.~\ref{fig4}(a)], the wave-packet size $a_{\rm wp}$ increases approximately linearly with $a_0$ [Fig.~\ref{fig4}(c)]. Consequently, the ratio $\langle z\rangle/a_{\rm wp}$ increases with $a_0$ [Fig.~\ref{fig4}(d)], slowly at small $a_0$ and faster at larger $a_0$, and becomes larger than unity at approximately $a_0\gtrsim 0.15$ ($\omega=3$), or $a_0\gtrsim 0.17$ ($\omega=5$), which is in accordance with the CMT peak position in Fig.~\ref{fig3}(c). Thus, the CMT increases with $a_0$ in the regime when the longitudinal displacement of the electron wave packet is larger than its spatial extent, and this behavior of CMT is reverted in the opposite case.

We will call the sub-regime of the interaction corresponding to the left part of Fig.~\ref{fig3}(c) the Coulomb-influenced regime, while the right part is referred to as the laser-dominated regime. To further justify this distinction, in Fig.~\ref{fig4}(b) we compare the explicit quantum expectation value of the electron displacement along the propagation direction at the end of the interaction, $\langle z\rangle$ with the corresponding classical Coulomb-free longitudinal drift distance $z_{cl}$, estimated as
\begin{equation}
z_{cl}=\int_{-\infty}^{\infty}\frac{A^2(t)}{2c}dt
\approx \frac{3}{8}\frac{a_0^2}{4}c\tau,
\label{eq:z0}
\end{equation}
where the factor $3/8$, in deviation from the plane-wave case~\cite{Salamin_1996}, arises from the integration with the explicit pulse shape. 
The relative deviation $ |z_{cl}-\langle z\rangle|/z_{cl}$ is shown in Fig.~\ref{fig4}(b).
It decreases systematically with increasing $a_0$ for both frequencies, demonstrating that the quantum longitudinal displacement progressively approaches the Coulomb-free classical drift prediction: $\langle z\rangle\approx z_{cl}$, at larger $a_0$. However, at $a_0\ll 1$, $\langle z\rangle$ is significantly smaller than $z_{cl}$, approaching the condition $ |z_{cl}-\langle z\rangle|/z_{cl}\sim 0.3$ at the transition region of the sub-regimes.

\begin{figure}
    \centering
    \includegraphics[width=1\linewidth]{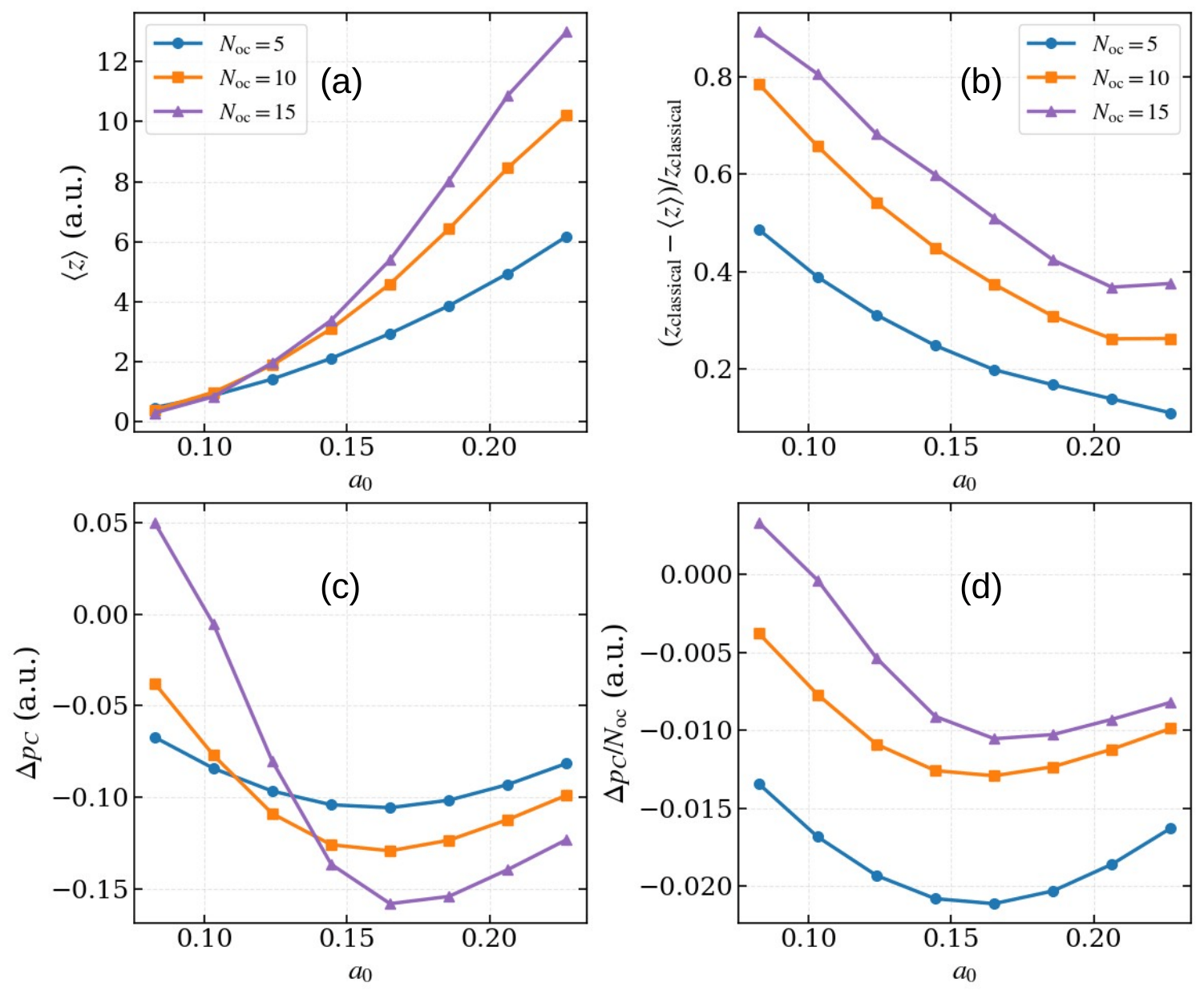}
    \caption{(a) Expectation value of the longitudinal electron displacement, $\langle z\rangle$, as a function of the pulse duration, with $N_c=5$, $10$, and $15$ optical cycles; (b) Relative deviation between the classical and quantum longitudinal displacements, $|z_{\mathrm{cl}}-\langle z\rangle|/z_{\mathrm{cl}}$; (c) The CMT along the propagation direction; (d) The CMT per optical cycle.}
    \label{fig5}
\end{figure}

\section{\label{sec:pulse}Pulse-duration effect in the sub-regimes}\label{sec:pulse}

In this section we analyze the effect of the pulse duration on the two sub-regimes. Does the pulse duration change the sub-regime, from the Coulomb-influenced to the laser-dominated? What are the specific features of the PMD when increasing the pulse duration, and how they depend on the sub-regime? 

We analyze these questions in the case of a laser field with a  frequency  of $\omega=3$~a.u. The nondipole TDSE is solved for pulse durations of $N_c=5$, $10$, and $15$ optical cycles, over a range of $a_0$ up to 0.25. To have an idea about the interaction sub-regime, the Coulomb-influenced  or the laser-dominated one, we have calculated the characteristic variables introduced in the previous section: the CMT in the laser propagation direction $\Delta p_C$, and the relative longitudinal displacement $|z_{\mathrm{cl}}-\langle z\rangle|/z_{\mathrm{cl}}$, see the results in Fig.~\ref{fig5}.
 
\begin{figure}[b]
    \centering
    \includegraphics[width=1\linewidth]{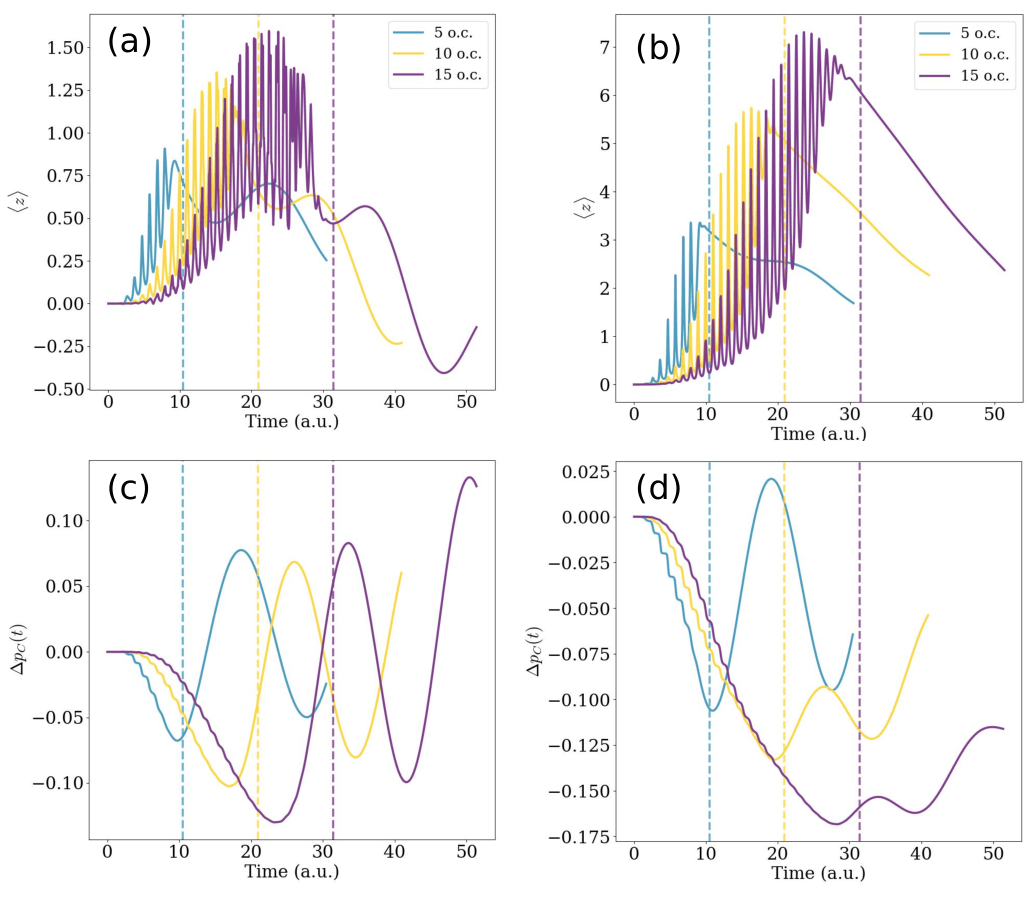}
    \caption{(a,b) Time evolution of the displacement expectation value $\langle z(t)\rangle$. (c,d) Corresponding CMT. The left column corresponds to the Coulomb-dominated regime ($E_0=40$~a.u.), while the right column  to the laser-dominated regime ($E_0=70$~a.u.).   }
    \label{fig6}
\end{figure}
 \begin{figure*}
    \centering
    \includegraphics[width=1\linewidth]{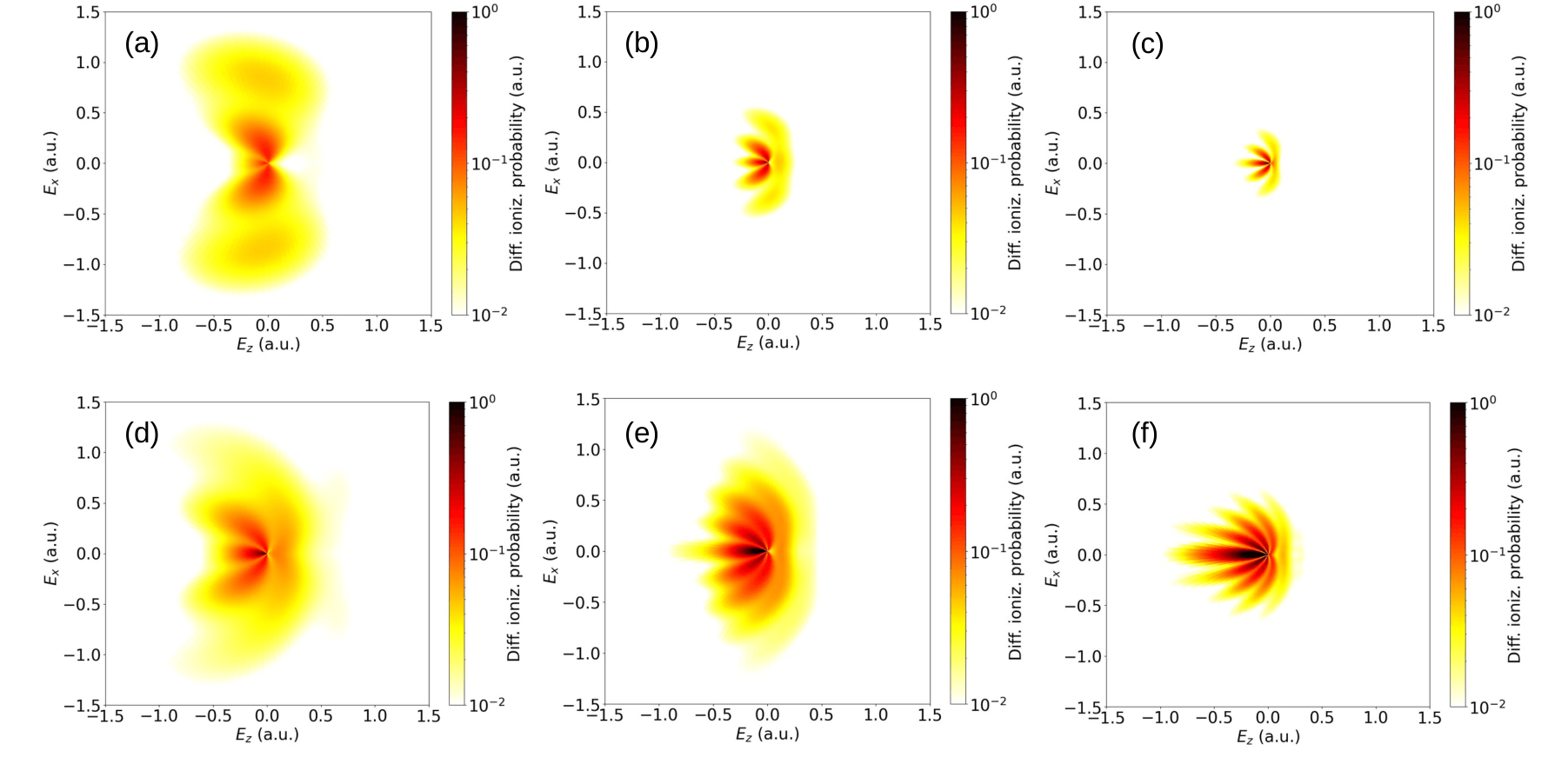}
    \caption{ZES for a hydrogen atom interacting with a  laser field with $\omega=3$~a.u. The laser field is $E_0=40$ ($a_0=0.097$) (top row); $E_0=70$ ($a_0=0.17$) (bottom row). The laser pulse duration in the number of  optical cycles is  $N_c=5$ (first column),  $N_c=10$ (middle column), and $N_c=15$ (last column).}
    \label{fig7}
\end{figure*}

The value of the longitudinal displacement $\langle z\rangle$, as shown in Fig.~\ref{fig5}(a), increases substantially with increasing pulse duration, as expected from the longer interaction time. At the same time, the relative deviation from the Coulomb-free classical displacement, shown in Fig.~\ref{fig5}(b), decreases with increasing $a_0$ for all three pulse durations, indicating a gradual transition toward laser-dominated dynamics. At low $a_0$ the Coulomb field  obviously hinders the drift, but this effect is suppressed at higher $a_0$. Although the magnitude of the displacement and its deviation from the classical prediction depend on the pulse duration, their overall dependence on $a_0$ remains qualitatively similar [cf.  with Fig.~\ref{fig4}(b)]. It is also instructive to see the time-resolved picture of the longitudinal displacement, presented in Fig.~\ref{fig6}(a,b) for two regimes. In the laser-dominated regime [Fig.~\ref{fig6}(b)], the Coulomb effect is negligible, and consequently, the forward displacement $\max\langle z(t)\rangle$ increases with pulse duration due to the accumulation of the laser-induced drift during the interaction. In contrast, in the Coulomb-influenced regime [Fig.~\ref{fig6}(a)], the Coulomb force partially compensates the laser-driven forward drift, and   consequently, the maximum displacement $\max\langle z(t)\rangle\sim 1.5$~a.u. is much smaller than the maximum of the Coulomb-free classical drift distance $z_0\sim 3.8$~a.u. and  remains nearly constant of the order of atomic size with the increase of the pulse duration.

For the accumulated CMT in Fig.~\ref{fig5}(c), we firstly note that it shows a minimum for all pulse durations similar to Fig.~\ref{fig3}(c), a property indicative of two sub-regimes. The value of $a_0$ corresponding to the minimum (transition between the sub-regimes) is not the same for different pulse durations in Fig.~\ref{fig5}(c), although the difference is not large. However, it is remarkable that the CMT per cycle  
\begin{eqnarray}
\eta_C=\Delta p_C/N_c,
\label{etaC}
\end{eqnarray}
shows regular behavior [Fig.~\ref{fig5}(d)] with the same minimum value for $a_0$ (approximately  at $a_0\approx 0.17$ for all pulse durations) [cf.  with Fig.~\ref{fig3}(c)]. Therefore, the parameter $\eta_C$ is well suitable for the classification of the interaction sub-regime in long laser pulses, and the sub-regime of the interaction defined in this way does not depend on the pulse duration. 

The pulse duration in Fig.~\ref{fig5}(c) affects the accumulated CMT differently in the two sub-regime. In the laser-dominated regime,  the longer laser pulse, the larger is the CMT (absolute value) to the electron. However, in the Coulomb-influenced regime, this behavior of CMT is reverted at rather low $a_0$, even yielding the positive  $\Delta p_C$. This could be related to the slow Coulomb orbiting during the interaction pointed out in Ref.~\cite{Boitsov_2026}. The latter feature, in particular, is illustrated in the time-resolved CMT calculations [Fig.~\ref{fig6}(c)], where during the interaction the CMT shows oscillating behavior (including positive values of CMT) in the Coulomb-influenced regime, but mostly negative increasing CMT in the laser-dominated one [Fig.~\ref{fig6}(d)].

 We have seen that the parameter defined by Eq.~(\ref{etaC}) well describes the sub-regime of the interaction in long laser pulses. However, it is not easy to estimate the CMT with given interaction parameters. Taking into account the results for the average longitudinal laser-induced drift and its time-resolved analysis [Fig.~\ref{fig6}(a,b)], we suggest a more simple dimensionless parameter characterizing the sub-regime: 
 %To quantify the role of the Coulomb interaction in the strong-field regime and the corresponding sub-regimes, we introduce the dimensionless parameter defined as 
 the ratio between the average longitudinal laser-induced drift per optical cycle and the characteristic atomic size:
\begin{equation}
\tilde{\eta}_C=\frac{z_0}{N_c a_s}\approx \frac{3}{8}\frac{ \lambda a_0^2}{4a_s},
\label{eq:zc}
\end{equation}
with the laser wavelength $\lambda=2\pi c/\omega$. Thus, $\tilde{\eta}_C$ governs the cycle-resolved competition between the laser-driven drift and the Coulomb interaction.
% whereas $z_0$ controls the overall interference pattern in the final angular distribution. 
At weak fields ($\tilde{\eta} \lesssim 1$) there is a strong influence of the Coulomb field on the drift, while at strong fields ($\eta_C \gg 1$) the drift during the interaction is almost not disturbed by the Coulomb field of the atomic core. The first case is the Coulomb-influenced  regime, and the second case -- is the laser dominated sub-regime. In the case of $\omega=3$ and $a_s=1$, the condition $\tilde{\eta} =1$ yields $a_0\approx 0.19$, which is a reasonable rough estimate [cf. Fig.~\ref{fig5}(d).

The dependence of the PMD pattern on the laser pulse duration in different sub-regimes is illustrated
in Fig.~\ref{fig7} (we concentrate on the ZES analysis). Two representative field strengths from the two sub-regimes are selected: $a_0=0.097$ ($E_0=40$~a.u.)  in the Coulomb-influenced regime, and $a_0=0.17$ ($E_0=70$~a.u.) in the laser-dominated one.
%corresponding to a laser intensity of $I=5.6\times10^{19}$~W/cm$^2$) in the Coulomb-influenced regime and $a_0=0.17$ ($E_0=70$~a.u., corresponding to $I=1.7\times10^{20}$~W/cm$^2$) in the laser-dominated regime.
The ZES in Fig.~\ref{fig7} exhibits markedly different behavior with the pulse duration increase in the different sub-regimes.
% at $E_0 = 40$ ~a.u., compared to the $E_0 = 70$~a.u. case. 
For $E_0 = 40$, a pronounced three-lobe structure appears at short pulse duration in Fig.6(a), shifted toward the counterpropagation direction.
As the pulse duration increases, the energies of the side lobes gradually decrease, while the energy of the main lobe remains nearly unchanged. In contrast, for $E_0 = 70$, the picture more familiar from previous studies emerges~\cite{Forre_2006,Zhou_2013,Telnov_2020,Telnov_Chu_2021,Geng_2021}, when the energy of the main lobe increases continuously with increasing pulse duration, and more additional side lobes emerge.

\begin{figure}
    \centering
    \includegraphics[width=0.8\linewidth]{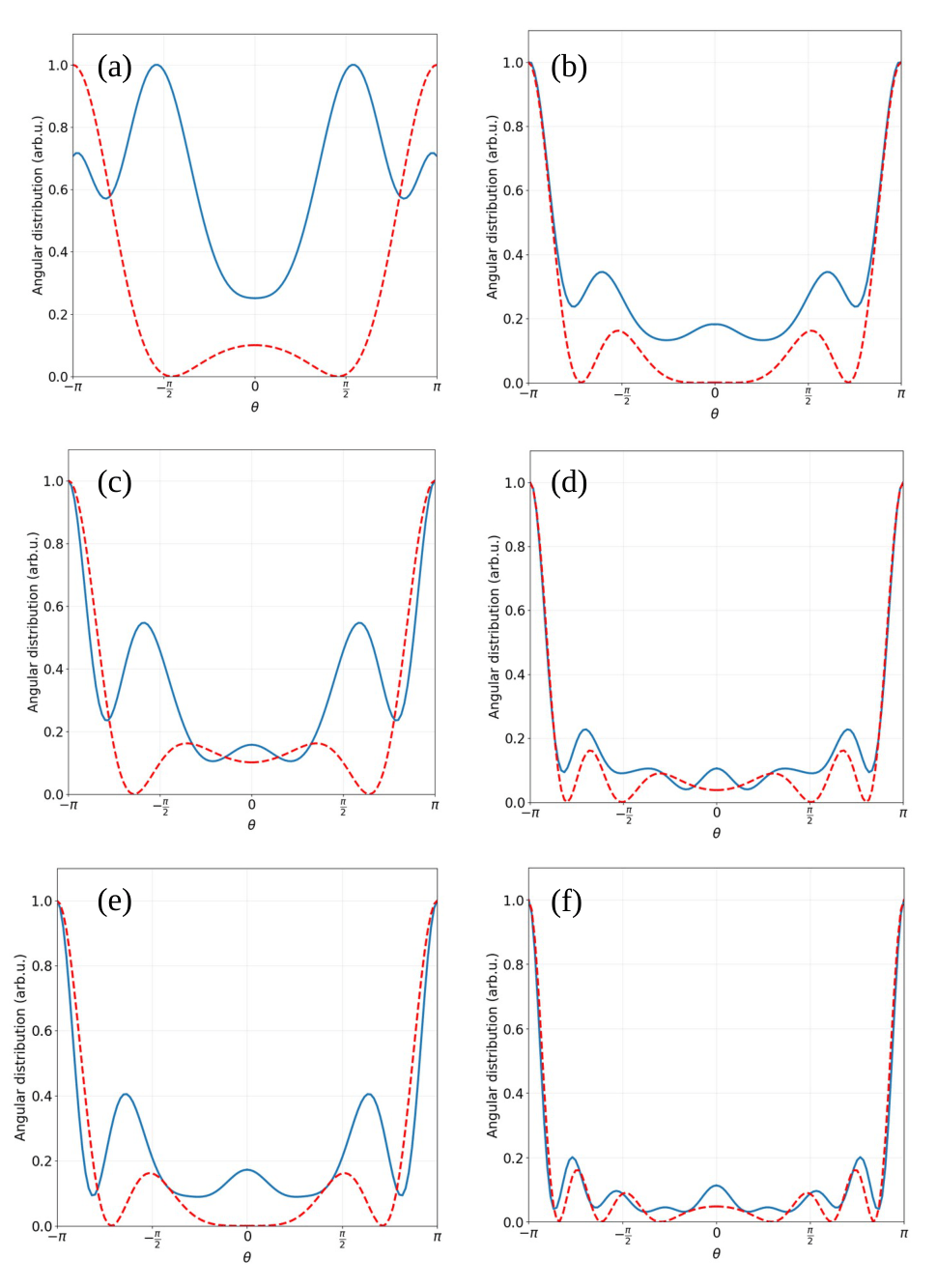}
    \caption{Angular distributions obtained from the analytical model of Eq.~(\ref{eq:angular}) (red dashed) and numerical simulations (blue solid). The left and right columns correspond to the Coulomb-dominated  ($E_0=40$~a.u., $a_0=0.097$) and laser-dominated  ($E_0=70$~a.u., $a_0=0.17$) regimes, respectively; the top, middle, and bottom rows correspond to pulse durations of $N_c=5$, $10$, and $15$ optical cycles, respectively.}
    \label{fig8}
\end{figure}

In the laser-dominated regime, the Coulomb effect during the interaction is assumed to be negligible. Therefore, the corresponding final angular distribution, see  Fig.~\ref{fig7}(d,e,f), can be interpreted within the analytical framework of Ref.~\cite{Geng_2021}, where the PMD structure is explained by the Fresnel diffraction of the ionized wave packet at the Coulomb center, and the interference pattern is governed by the total Coulomb-free displacement $z_0$.  According to Ref.~\cite{Geng_2021}, the angular distribution of ZES can be expressed as:
\begin{equation}
|a(\theta, z_0)|^2 \propto
\left|
\sum_{l=0}^{l_{\max}}
\frac{2l+1}{\sqrt{z_0}}
J_{2l+1}\!\left(\sqrt{8z_0}\right)\, P_l(\cos\theta)
\right|^2,
\label{eq:angular}
\end{equation}
where $J_{2l+1}$ are the Bessel functions, and $P_l$  Legendre polynomials. Let us compare the analytical description of the angular distribution of Eq.~(\ref{eq:angular}) with out simulations. In Fig.~\ref{fig8}, the red dashed lines represent the analytical results based on Eq.~(\ref{eq:angular}), while the blue solid lines correspond to the numerical simulations. We see that the Coulomb-free description is relevant in the laser-dominated regime, but it failed in the Coulomb-influenced  one, justifying the names of the regimes.

The analytical model of Ref.~\cite{Geng_2021} assumes that the emitted electron can be treated as
originating from an effective point-like source located far from the atomic core. This approximation is valid only when the electron wave packet is rapidly driven far away from the nucleus by a sufficiently strong laser field.
However, for $E_0=40$ the electron remains close to the ionic core for a
considerable time after ionization. As a consequence, the Coulomb potential
cannot be neglected and the electron cannot be approximated as a free
point-like emitter.
The long-range Coulomb interaction significantly modifies the forward
scattering dynamics, which is essential for the formation of the zero-angle
peak. Since this effect is not fully captured by the simplified theoretical
model, the zero-angle peak is not reproduced in Fig.~\ref{fig8}(a,c,e).\\

\section{Conclusions}\label{sec:concl}

We have studied the ionization of a hydrogen atom in  strong linearly polarized x-ray fields via numerical solutions of the nondipole TDSE and analyzed the Coulomb field effects for the near zero-energy structure.

Our analysis  shows that the evolution of the ZES is governed by the interplay between the laser-driven longitudinal drift and the Coulomb-induced momentum transfer. We have identified two sub-regimes of the interaction. Increasing the field nondipole parameter $a_0$,  the dynamics exhibits a transition from a Coulomb--laser competing (Coulomb-influenced) regime  to a radiation-pressure-dominated (laser-dominated) regime. In the former case, the Coulomb momentum transfer is comparable to the laser-induced drift momentum and partially compensates it, leading to a nearly constant longitudinal displacement and stable main-lobe energy with increasing pulse duration. In the latter case, the drift momentum dominates, resulting in a continuous increase of the forward displacement and a corresponding growth of the main-lobe energy with increasing the pulse duration. A distinct criterion of the regime is provided by the CMT dependence on the laser field strong-field parameter $a_0$. The absolute value of CMT increases with $a_0$ in the Coulomb-influenced  regime, while this dependence is opposite in the laser-dominated regime. The transition between the regimes takes place at $a_0$ corresponding to the maximum of the absolute value of CMT. Our simulations show that the PMD pattern evolution across the regimes is essentially dependent on the laser period and the laser pulse duration.

Overall, our results provide clear criteria for identifying the dominant mechanism and typical features of the Coulomb effect governing photoelectron momentum distributions beyond the dipole approximation.

\bibliography{strong_fields_bibliography}

\appendix

 \end{document}